\documentclass[aps,prb,twocolumn,groupedaddress,showpacs]{revtex4} 

\usepackage{graphicx}
\usepackage{dcolumn}
\usepackage{bm}

\begin{document} 

\title{Thermal conductivity of lightly Sr- and Zn-doped La$_2$CuO$_4$ 
single crystals} 

\author{X. F. Sun} 
\email[]{ko-xfsun@criepi.denken.or.jp} 
\author{J. Takeya} 
\author{Seiki Komiya} 
\author{Yoichi Ando} 
\email[]{ando@criepi.denken.or.jp} 
\affiliation{Central Research Institute of Electric Power 
Industry, Komae, Tokyo 201-8511, Japan.} 

\date{\today} 

\begin{abstract} 
Both $ab$-plane and $c$-axis thermal conductivities 
($\kappa_{ab}$ and $\kappa_c$) of lightly doped 
La$_{2-x}$Sr$_x$CuO$_4$ and La$_2$Cu$_{1-y}$Zn$_y$O$_4$ single 
crystals ($x$ or $y$ = 0 -- 0.04) are measured from 2 to 300 K. 
It is found that the low-temperature phonon peak (at 20 -- 25 K) 
is significantly suppressed upon Sr or Zn doping even at very low 
doping, though its precise doping dependences show interesting 
differences between the Sr and Zn dopants, or between the $ab$ 
plane and the $c$ axis. Most notably, the phonon peak in 
$\kappa_c$ decreases much more quickly with Sr doping than with 
Zn doping, while the phonon-peak suppression in $\kappa_{ab}$ 
shows an opposite trend. It is discussed that the scattering of 
phonons by stripes is playing an important role in the damping of 
the phonon heat transport in lightly doped LSCO, in which static 
spin stripes has been observed by neutron scattering. We also 
show $\kappa_{ab}$ and $\kappa_c$ data of 
La$_{1.28}$Nd$_{0.6}$Sr$_{0.12}$CuO$_4$ and 
La$_{1.68}$Eu$_{0.2}$Sr$_{0.12}$CuO$_4$ single crystals to 
compare with the data of the lightly doped crystals for the 
discussion of the role of stripes. At high temperature, the 
magnon peak (i.e., the peak caused by the spin heat transport 
near the N\'{e}el temperature) in $\kappa_{ab}(T)$ is found to be 
rather robust against Zn doping, while it completely disappears 
with only 1\% of Sr doping. 
\end{abstract} 

\pacs{74.25.Fy, 74.62.Dh, 74.72.Dn} 

\maketitle 

\section{Introduction} 

It has recently been discussed that the holes in the high-$T_c$ 
cuprates self-organize into quasi-one-dimensional 
stripes.\cite{Tranquada, Yamada, Mook1, Mook2, Wakimoto, 
Matsuda1, Matsuda2, Fujita, Hunt, Ando1, Noda, Zhou, Emery, 
Kivelson1, Carlson, Zaanen, Ando2} The stripe phase is a periodic 
distribution of antiferromagnetically-ordered spin regions 
separated by quasi-one-dimensional charged domain walls which act 
as magnetic antiphase boundaries. Although the relation between 
the stripe correlations and the mechanism of high-$T_c$ 
superconductivity is not fully understood yet, it has become 
clear \cite{Ando1,Ando2} that the charge stripes determine the 
charge transport behavior, at least in the lightly hole doped 
region: charges can move more easily along the stripes than 
across the stripes.\cite{Ando2} Given that the stripes indeed 
affect the basic physical properties such as charge transport, it 
is desirable to build a comprehensive picture of the roles of 
stripes in the cuprates. Since the nonuniform charge distribution 
is expected to induce variations of the local crystal 
structure,\cite{Tranquada,Mook2} which disturb phonons, the 
phonon heat transport is expected to be a good tool capable of 
detecting the influence of stripes even in the charge-localized 
region. 

Thermal conductivity is one of the basic transport properties that 
provides a wealth of useful information on the charge carriers and 
phonons, as well as their scattering processes. 
It is known that the antiferromagnetic (AF) insulating compound 
La$_2$CuO$_4$ shows predominant phonon transport at low temperatures, 
which is manifested in a large phonon peak at 20 -- 25 K in the 
temperature dependence of both $ab$-plane and $c$-axis thermal 
conductivities ($\kappa_{ab}$ and $\kappa_c$);\cite{Nakamura} such 
phonon peak disappears in Sr-doped La$_{2-x}$Sr$_x$CuO$_4$ (LSCO) 
with $x$ = 0.10 -- 0.20.\cite{Nakamura} 
The suppression of the phonon peak is normally caused by the 
defect scattering and the electron scattering of phonons in 
doped single crystals; 
however, it is also known that the phonon peak re-appears in both 
$\kappa_{ab}(T)$ and $\kappa_c(T)$ of overdoped 
La$_{1.7}$Sr$_{0.3}$CuO$_4$,\cite{Nakamura} which cannot be 
explained in this scenario. 
Furthermore, it was found that in rare earth (RE) and Sr co-doped 
La$_2$CuO$_4$, such as La$_{1.28}$Nd$_{0.6}$Sr$_{0.12}$CuO$_4$, 
the phonon thermal conductivity is much more enhanced in the 
non-superconducting LTT (low-temperature tetragonal) phase, 
compared to that in LSCO with the same Sr content.\cite{Baberski} 
This cannot happen if the defect scattering and electron scattering 
of phonons are the only source of the peak suppression. 
Based on the fact that in RE- and Sr-doped La$_2$CuO$_4$ 
systems the phononic thermal conductivity is always strongly 
suppressed in superconducting samples,\cite{Baberski} 
it was proposed that dynamical stripes cause a pronounced 
damping of phonon heat transport, while static stripes do not 
suppress the phonon transport so significantly.\cite{Baberski} 
It is indeed possible that the static stripes are not effective in 
scattering phonons, if they only induce {\it periodic} local 
distortions in the crystal structure. 
Since the spin stripes in lightly doped LSCO ($x$ = 0.01 -- 0.05) are 
reported to be static,\cite{Wakimoto, Matsuda1, Matsuda2, Fujita} 
one may naively expect that the phonon heat transport is not strongly 
suppressed in such lightly doped LSCO. 
However, there has been no measurement of the thermal conductivity 
of lightly doped LSCO single crystals. 

\begin{figure*} 
\includegraphics[clip,width=13.5cm]{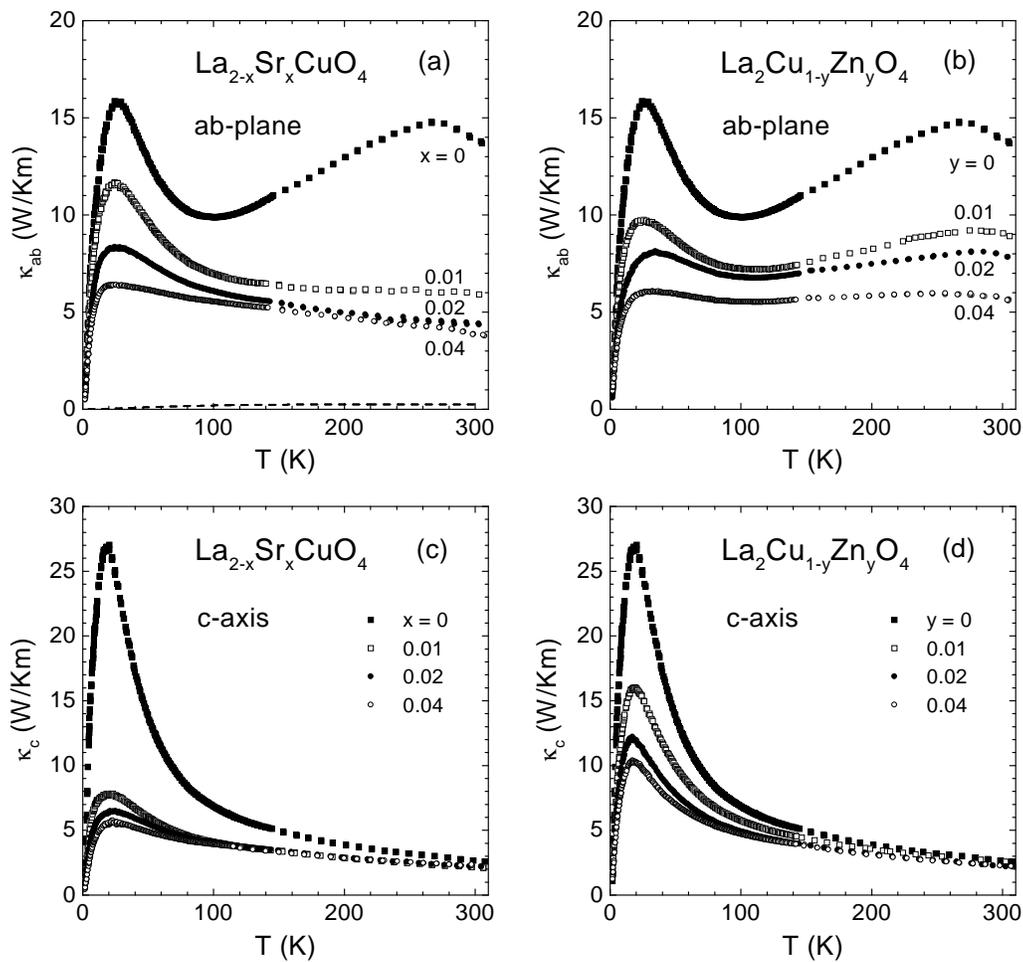} 
\caption{Thermal conductivity of lightly doped 
La$_{2-x}$Sr$_x$CuO$_4$ and La$_2$Cu$_{1-y}$Zn$_y$O$_4$ single 
crystals along the $ab$ plane and the $c$ axis. Dashed line in 
panel (a) is $\kappa_{e,ab}$ of La$_{1.96}$Sr$_{0.04}$CuO$_4$ 
estimated from the Wiedemann-Franz law.} 
\end{figure*} 

In this paper, we report our study of the thermal conductivity of 
lightly doped La$_{2-x}$Sr$_x$CuO$_4$ ($x$ = 0 -- 0.04) single crystals. 
Such low doping levels are intentionally selected for the reasons of 
both tracing the evolution of the phonon peak and avoiding significant 
modifications of the crystal structure and phonon mode, which may 
complicate interpretations of the data. It is found that the 
phonon peak is suppressed significantly even with very small 
doping concentration, especially for the $c$-axis heat transport. 
Since this result is contrary to the naive expectation mentioned in 
the previous paragraph, we also study the thermal conductivity of 
La$_2$Cu$_{1-y}$Zn$_y$O$_4$ (LCZO) ($y$ = 0 -- 0.04) single 
crystals, which have similar amount of dopants as LSCO. 
The important difference between these two systems is that there 
cannot be stripes in LCZO (since there is no carrier), 
while there are static spin stripes in LSCO at low temperatures. 
By comparing the thermal conductivity behaviors in these two systems, 
we can elucidate whether the static stripes are responsible for 
strong scattering of phonons. 
The comparison indicate that the stripes, though they are static, 
indeed damp the $c$-axis phonon transport significantly, while their 
role is minor in the in-plane phonon transport. 

\section{Experiments} 

The single crystals of LSCO and LCZO are grown by the 
traveling-solvent floating-zone (TSFZ) technique and carefully 
annealed in flowing pure He gas to remove the excess 
oxygen.\cite{Komiya} 
After the crystallographic axes are determined 
by using the X-ray Laue 
analysis, the crystals are cut into rectangular thin platelets 
with the typical sizes of $2.5 \times 0.5 \times 0.15$ mm$^3$, 
where the $c$ axis is perpendicular or parallel to the platelet 
with an accuracy of 1$^{\circ}$. 
The thermal conductivity $\kappa$ is measured in the 
temperature range of 2 -- 300 K using a 
steady-state technique;\cite{AndoCuGeO,Ando3,Takeya} 
above 150 K, a double thermal shielding is employed to minimize 
the heat loss due to radiation, and the residual radiation loss is 
corrected for by using an elaborate measurement 
configuration. 
The temperature difference $\Delta$$T$ in the sample is measured 
by a differential Chromel-Constantan thermocouple, which is glued 
to the sample using GE vanish. 
The $\Delta$$T$ varies between 0.5\% and 2\% of the 
sample temperature. To improve the accuracy of the measurement at 
low temperatures, $\kappa$ is also measured with 
``one heater, two thermometer" method from 2 to 20 K by using a 
chip heater and two Cernox chip sensors.\cite{AndoCuGeO} 
The errors in the thermal conductivity data are smaller than 10\%, 
which is mainly caused by the uncertainties in the geometrical factors. 
Magnetization measurements are carried out using a 
Quantum Design SQUID magnetometer. 

\section{Results} 

\subsection{Anisotropic Heat transport in La$_2$CuO$_4$} 

It is useful to first establish an understanding of the anisotropic 
heat transport in undoped crystals. The temperature dependences of the 
thermal conductivity measured along the $ab$ plane and the $c$ axis in 
pure La$_2$CuO$_4$ are shown in Fig. 1, which also contains data for 
LSCO and LCZO single crystals. For undoped La$_2$CuO$_4$ sample, a 
sharp peak appears at low temperature, at $\sim$25 K in 
$\kappa_{ab}(T)$ and at $\sim$20 K in $\kappa_c(T)$, respectively. In 
the $c$ direction, above the peak temperature, $\kappa_c$ decreases 
with increasing $T$ approximately following $1/T$ dependence, which is 
typical for phonon heat transport.\cite{Berman} This phonon peak 
originates from the competition between the increase in the population 
of phonons and the decrease in their mean free path (due to the 
phonon-phonon umklapp scattering) with increasing temperature. In 
contrast to the mainly phononic heat transport in the $c$ direction, 
in the $ab$ plane another large and broad peak develops at higher 
temperature ($\sim$270 K), which has been attributed \cite{Nakamura} 
to the magnon transport in a long-range-ordered AF state. The heat 
conduction due to magnetic excitations has been observed in many 
low-dimensional quantum antiferromagnets, such as one-dimensional spin 
systems CuGeO$_3$,\cite{Takeya,AndoCuGeO,Vasilev} 
Sr$_2$CuO$_3$,\cite{Sologubenko1} SrCuO$_2$,\cite{Sologubenko2} and 
(Sr,Ca)$_{14}$Cu$_{24}$O$_{41}$,\cite{Sologubenko3} and also 
two-dimensional antiferromagnet K$_2$V$_3$O$_8$.\cite{Sales} The 
absence of this high-$T$ peak in $\kappa_c(T)$ is obviously due to the 
much weaker magnetic correlations in the $c$ direction compared to 
that in the CuO$_2$ plane. All these behaviors in undoped 
La$_2$CuO$_4$ are consistent with the previously reported 
data.\cite{Nakamura} It is worthwhile to note that the phononic peak 
value of $\kappa_c$ (27 W/Km) is considerably larger than that of 
$\kappa_{ab}$ (16 W/Km). One possible reason for this difference is 
that the phonon heat transport is intrinsically easier along the $c$ 
axis, which is not so easy to conceive in view of the layered crystal 
structure of La$_2$CuO$_4$. Another, more likely possibility is that 
there exists phonon-magnon scattering in the $ab$ plane that causes 
additional damping of the phonon peak. Such scattering may also happen 
in the $c$ axis, however, it should be much weaker than in the $ab$ 
plane because it is well known that the magnons are good excitations 
only for the in-plane physics in the La$_2$CuO$_4$ system. 

\subsection{Doping dependence of magnon peak and N\'{e}el 
transition} 

\begin{figure} 
\includegraphics[clip,width=6cm]{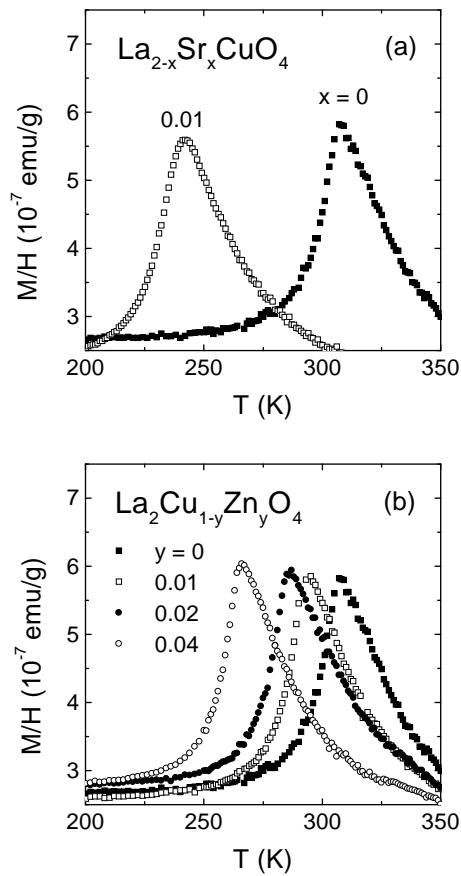} 
\caption{Magnetic susceptibility of (a) La$_{2-x}$Sr$_x$CuO$_4$ 
and (b) La$_2$Cu$_{1-y}$Zn$_y$O$_4$ single crystals measured in 
0.5 T field applied along the $c$ axis.} 
\end{figure} 

Upon Sr or Zn doping, there are strong doping dependences in both 
$\kappa_{ab}$ and $\kappa_c$. Let us first discuss the change of 
the high-$T$ magnon peak in $\kappa_{ab}(T)$ for different 
dopants. This peak is suppressed very quickly upon Sr-doping and 
in fact completely disappears at only 1\% doping concentration. 
On the other hand, the suppression of this peak is much slower in 
the Zn-doped case, where the peak, though gradually suppressed, 
survives to $y$ = 0.04. To clarify the relation between the 
high-$T$ peak and the N\'{e}el order, the magnetic susceptibility 
of LSCO and LCZO is measured from 5 to 350 K in the magnetic 
field of 0.5 T applied along the $c$ axis. Figure 2 shows the 
temperature dependences of the magnetic susceptibility, where the 
peak corresponds to the N\'{e}el transition (no transition is 
observed in LSCO with $x \ge$ 0.02). It is clear (as has been 
already reported \cite{Chou,Keimer}) that Sr doping is much 
stronger than Zn doping in destroying the AF long-range order, 
where the former decreases the N\'{e}el temperature $T_N$ much 
more quickly. The reason is related to the fact that holes 
introduced by Sr are mobile and have 1/2 spin, while Zn (zero 
spin) brings static spin vacancy in CuO$_2$ plane. Figure 3 
summarizes the doping dependences of the size of the high-$T$ 
peak in $\kappa_{ab}(T)$ [defined by the difference between the 
peak value and $\kappa_{ab}$(100 K)], as well as those of the 
N\'{e}el temperature $T_N$, for the two systems. This result 
confirms that the magnon peak is quickly diminished as $T_N$ is 
reduced. It is useful to note that the disorder in the N\'{e}el 
state induced by Sr doping appears to be detrimental to the 
magnon heat transport, because the magnon peak completely 
disappears at $x$ = 0.01 even though the N\'{e}el order is still 
established at 240 K for this Sr doping. 

\begin{figure} 
\includegraphics[clip,width=5.5cm]{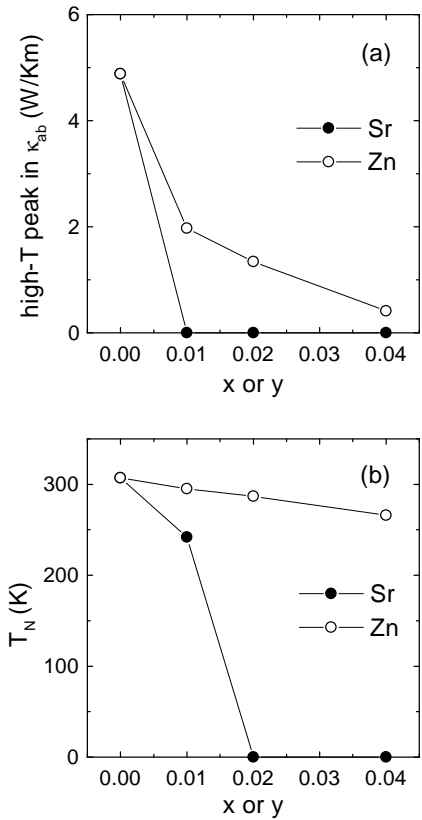} 
\caption{(a) Doping dependences of the size of high-$T$ peak in 
$\kappa_{ab}(T)$ [defined by the difference between the peak 
value and $\kappa_{ab}$(100 K)] of La$_{2-x}$Sr$_x$CuO$_4$ and 
La$_2$Cu$_{1-y}$Zn$_y$O$_4$ single crystals. (b) Doping 
dependences of the N\'{e}el temperature $T_N$ for the two cases.} 
\end{figure} 

\subsection{Doping dependence of phonon peak} 

More interesting changes happen in the suppression of the phonon 
peak at low temperatures. One can see in Fig. 1 that in 
$\kappa_{ab}(T)$ the peak magnitude decreases more quickly with 
Zn doping than with Sr doping; on the contrary, the peak value in 
$\kappa_c$ decreases much more quickly with Sr doping than with 
Zn doping.  It should be noted that those peculiar differences 
between Sr doping and Zn doping cannot be due to the additional 
electronic thermal conductivity in the Sr-doped samples: The 
electronic thermal conductivity $\kappa_e$ can roughly be 
estimated by the Wiedemann-Franz law, $\kappa_e=L_0T/\rho$, where 
$\rho$ is the electrical resistivity and $L_0$ is the Lorenz 
number (which can be approximated by the Sommerfeld value, $2.44 
\times 10^{-8}$ W$\Omega$/K$^2$). Using the resistivity data for 
our crystals,\cite{Ando5} the contribution $\kappa_{e,ab}$ for 
La$_{1.96}$Sr$_{0.04}$CuO$_4$ can be estimated to be smaller than 
0.25 W/Km in the whole temperature region [dashed line in Fig. 
1(a)], and samples with lower Sr doping should have smaller 
$\kappa_{e,ab}$ than this estimate for $x$ = 0.04; clearly, the 
electronic contribution can be safely neglected in the lightly Sr 
doped region in the discussion of the $\kappa_{ab}(T)$ behavior, 
not to mention the $\kappa_{c}(T)$ behavior. Thus, the changes in 
the low-temperature peak upon doping must be due to the changes 
in the phonon transport properties, that is, the scattering 
process that determines the phonon mean free path. 

\subsection{Re-appearance of the phonon peak in RE-doped LSCO} 

\begin{figure} 
\includegraphics[clip,width=6.5cm]{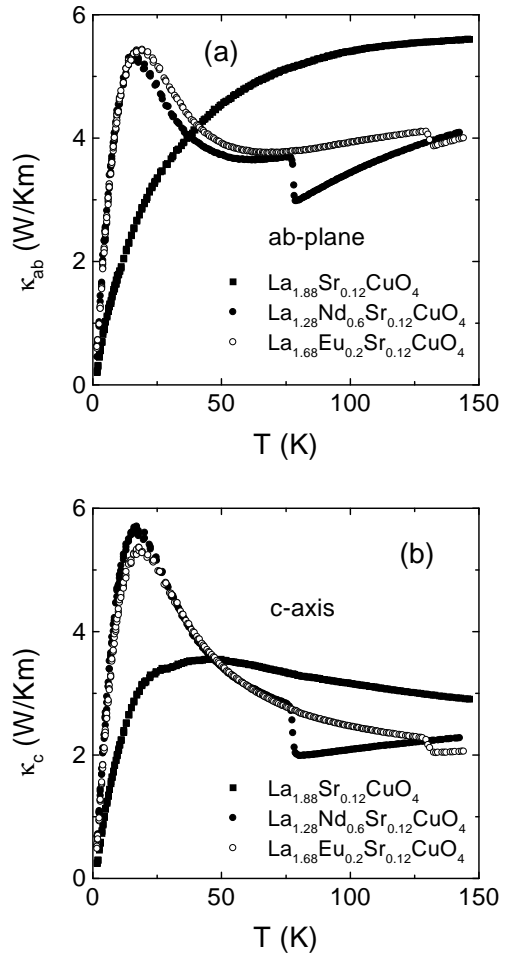} 
\caption{Comparison of the low-temperature thermal conductivity 
of LSCO ($x$ = 0.12) single crystals to that of Nd- and Eu-doped 
crystals, in which the phonon peak re-appears. 
The data for both (a) $\kappa_{ab}(T)$ and (b) $\kappa_{c}(T)$ 
are shown.} 
\end{figure} 

Although both the phonon-impurity scattering and the 
phonon-carrier scattering can contribute to the suppression of the 
phonon peak, these scattering processes cannot be the only mechanisms 
to determine the Sr-doping dependence of the phonon heat transport, 
because the phonon peak re-appears in overdoped LSCO and in RE-doped 
LSCO.\cite{Nakamura, Baberski} 
Figure 4 shows our data for the re-appearance of the phonon peak 
in RE-doped samples (both Nd and Eu doped cases) for $x$ = 0.12; 
these data are taken on single crystals, and essentially confirms 
the polycrystalline data reported by Baberski {\it et al.} 
\cite{Baberski}  The single crystal data of Fig. 4 allow us to 
compare the absolute values of $\kappa_{ab}$ and $\kappa_c$ of the 
RE-doped samples to those of the RE-free LSCO; such comparison tells us 
that the low-$T$ peak values of $\kappa_{ab}$ and $\kappa_c$ of the 
RE-doped crystals at $x$ = 0.12 are similar to those of the 
RE-free LSCO crystals at $x$ = 0.04, despite the factor of three 
difference in the Sr concentrations.  This result clearly indicates 
that the phonon-impurity scattering and the phonon-carrier scattering 
are not the only scattering mechanisms to determine the phonon 
mean free path. 

\section{Discussion} 

\subsection{Magnetic scattering of phonons in the $ab$ plane of 
LCZO} 

Based on the apparent similarity of the doping dependence of the phonon
peak in $\kappa_{ab}$, shown in Figs. 1(a) and 1(b), one might naively
conclude that the lattice disorders induced by Zn and Sr in the $ab$
plane are very similar and that the $\kappa_{ab}(T)$ behavior is
essentially explained only by the lattice disorder; however, this is not
likely to be the case, which can be understood by considering the nature
of the impurity scattering and the additional charge carriers introduced
in LSCO. First, it is important to notice that the low-$T$ phonon peak
in $\kappa_{ab}$ is suppressed slightly more quickly by Zn doping than
by Sr doping. In the impurity-scattering scenario,\cite{Berman} it is
difficult to conceive that the phonons are scattered more strongly in
LCZO than in LSCO, because the atomic mass difference between Cu and Zn
is much smaller than that between La and Sr. In addition, the
contribution of the phonon-carrier scattering, which exists only in
LSCO, would cause the phonon peak in LSCO to be suppressed more quickly
than in LCZO. Thus, the difference between the Sr-doping dependence and
the Zn-doping dependence in $\kappa_{ab}(T)$ at low temperature, which
is opposite to the naturally expected trend, suggests that there are
additional scatterers of phonons in the Zn doped samples. It is most
likely that the additional scatterers in LCZO are of magnetic origin;
magnons, which cause the high-temperature peak in LCZO, are an obvious
candidate, and the magnetic disordering around Zn atoms may also cause
some scattering of phonons through magnetoelastic
coupling.\cite{Cordero} It is useful to note that, as we mentioned in
section III A, magnons are likely to be already responsible for the
relatively small phonon peak in $\kappa_{ab}$ (compared to that in
$\kappa_c$) in pure La$_2$CuO$_4$.

\subsection{Scattering of phonons by static spin stripes in LSCO} 

Contrary to the relatively small difference in the phonon peak of
$\kappa_{ab}$ between LSCO and LCZO, the suppression of the phonon peak
of $\kappa_c$ is much quicker in LSCO than in LCZO; especially, the
sharp suppression of the phonon peak from $x$ = 0 to 0.01 is quite
surprising and is the most striking observation of this work. Apart 
from the impurity scattering, two different scattering processes of
phonons are possibly responsible for the suppression of the $c$-axis 
phonon heat transport. The first possibility is the magnon-phonon 
scattering. This may partly play a role (particularly in samples where 
the long-range N\'{e}el order is established) in the $c$-axis phonon 
heat transport, although their role is expected to be minor because of 
the essentially two-dimensional nature of the magnons in the LCO 
system, which means that the magnons cannot effectively change the 
wave vector of the $c$-axis phonons. The second possibility is the 
scattering of phonons caused by the static stripes that exist only in 
LSCO.  In the following, we elaborate on this possibility. 

It is known that static spin stripes are formed at low temperatures 
in lightly doped LSCO.\cite{Wakimoto, Matsuda1, Matsuda2, Fujita} 
If the charges also form static stripes together with the spins, 
they certainly cause local lattice distortions. 
Even if the charges do not conform to the static stripe 
potentials set by the spins, the spin stripes themselves 
may well cause local lattice distortions due to the magnetoelastic 
coupling;\cite{Cordero} this direct coupling between the spin stripes 
and the local lattice distortions is in fact very likely, because 
recently Lavrov {\it et al.} found that the spin-lattice coupling 
is very strong in lightly doped LSCO.\cite{Lavrov} 
It should be noted that the local lattice distortions due to static 
stripes are {\it not} expected to scatter phonons when 
the stripes induce only static and {\it periodic} modulation of the 
lattice (which is in fact a kind of superlattice); 
however, possible disordering of stripes in the $c$ direction 
can introduce rather strong scattering of phonons in this direction. 
In fact, because of the weak magnetic correlations along the $c$ axis, 
the stripes in neighboring CuO$_2$ planes are only weakly 
correlated or even uncorrelated, which is best evidenced by the 
very short magnetic correlation length $\xi_c$ in the stripe phase 
(usually smaller than the distance between neighboring CuO$_2$ 
planes) in the lightly doped LSCO.\cite{Matsuda1,Matsuda2} 

We can further discuss the possible role of static stripes in the
suppression of the phonon peak in $\kappa_{ab}$ of lightly doped 
LSCO. Although the stripes are much better ordered in the $ab$ 
plane than in the $c$ direction,\cite{Matsuda1, Matsuda2} there 
are two reasons that phonons in the $ab$ plane are possibly 
scattered by the stripes. First, the static spin stripes were 
reported to be established at about 30 K to 17 K for $x$ = 0.01 
to 0.04 by the neutron measurements.\cite{Wakimoto,Matsuda2} 
These temperatures are very close to the position of the phonon 
peak (20 -- 25 K). Near the stripe freezing temperature, local 
lattice distortions are expected to be well developed and yet are 
slowly fluctuating (or disordered), which would tend to scatter 
phonons. Second, there is no evidence until today that well 
periodically-ordered static charge stripes are formed in lightly 
doped LSCO; instead, our charge transport data strongly suggest 
\cite{Ando2,Ando5,Kivelson2} that the charge stripes exist in a 
liquid (or nematic) state.\cite{Kivelson1} Such disordered state 
of charge stripes would tend to scatter phonons. Thus, it is 
possible that the static stripes also contribute to the 
scattering of phonons in the in-plane heat transport in LSCO, 
although their effect appears to be much weaker than that in the 
$c$ direction. 

\section{Summary} 

We have measured the $ab$-plane and the $c$-axis thermal 
conductivities of lightly doped LSCO and LCZO single crystals. It 
is found that the low-temperature phonon peak is significantly 
suppressed upon Sr or Zn doping even at very low doping levels, 
and that the doping dependence show clear differences between 
the Sr and Zn dopants, and between $\kappa_{ab}$ and $\kappa_{c}$. 
The experimental observations can be summarized as follows:
(i) The phonon peak in $\kappa_c$ decreases much more 
quickly with Sr doping than with Zn doping.
(ii) On the other hand, the phonon peak 
in $\kappa_{ab}$ is suppressed slightly more quickly with Zn doping than 
with Sr doping. 
(iii) At high temperature, the magnon peak in $\kappa_{ab}(T)$ decreases 
much more quickly with Sr doping than with Zn doping; in fact, the magnon peak 
completely disappears in LSCO with $x$ = 0.01, while 
it is still observable in LCZO with $y$ = 0.04. 
(iv) Rare-earth (Nd or Eu) doping for LSCO ($x$ = 0.12) enhances 
the phonon heat transport in both $\kappa_{ab}$ and $\kappa_c$, 
and this is manifested in the re-appearance of the low temperature
phonon peak, whose height is similar to that of rare-earth-free LSCO 
with $x = 0.04$. 

Based on the peculiar doping dependences of the low-temperature 
phonon peak, we can deduce the following conclusions for the 
phonon heat transport at low temperatures: (i) In the in-plane 
direction, disordered static stripes are probably working as 
scatterers of phonon (in addition to the Sr impurities and the 
holes) in LSCO, while in LCZO magnetic scatterings cause strong 
damping of the phonon transport, which overcompensates the 
absence of the stripe scattering. (ii) In the $c$-axis heat 
transport, besides the possible contribution of the magnon-phonon 
scattering, strong disorder of the stripe arrangement along the 
$c$-axis is mainly responsible for the scattering of phonons. The 
latter scattering mechanism naturally explains the strong damping 
of the phonon peak in LSCO, while allowing much slower 
suppression of the phonon peak in LCZO where there is no stripe. 
Therefore, the effect of static spin stripes, which are formed in 
the lightly doped LSCO, appears to be most dramatically observed 
in the $c$-axis phonon heat transport.

\begin{acknowledgments} 
We thank A. N. Lavrov and I. Tsukada for helpful discussions. 
X.F.S. acknowledges support from JISTEC. 
\end{acknowledgments}

\end{document}